\newcommand{\tindex}[1]{{\scriptstyle {\rm#1}}}
\documentstyle[pre,aps,epsf,floats]{revtex}

\begin{document} 
\flushbottom
\draft

\twocolumn[\hsize\textwidth\columnwidth\hsize\csname @twocolumnfalse\endcsname
  
\title{%
  Dimensional crossover and driven interfaces in disordered ferromagnets
  } 
\author{
  L.\ Roters\cite{mail4lars} and K.\,D.\ Usadel\cite{mail4usadel}
  }
\address{
  Theoretische Tieftemperaturphysik, \\
  Gerhard-Mercator-Universit\"at Duisburg,\\ 
  47048 Duisburg, Germany \\
  }
\date\today
\maketitle
\begin{abstract}
  We study the depinning transition of driven interfaces in thin
  ferromagnetic films driven by external magnetic fields.
  Approaching the transition point the correlation length increases
  with decreasing driving.
  If the correlation length becomes of the order of the film thickness
  a crossover to two dimensional behavior occurs.
  From the corresponding scaling analysis we
  determine the exponents characterizing the transition of the three
  dimensional system. 
\end{abstract}
]
\narrowtext
\section{introduction}
\label{sec:intro}
We study interface motion in the random-field Ising model. 
In this model an interface separates regions of opposite spin
orientation.
A magnetic field forces the interface to move whereas this motion is
hindered by the disorder. 
At zero temperature a permanent interface motion is found if the
driving field $H$ exceeds a critical threshold $H_\tindex{c}$
determined by the disorder.
The vanishing of the interface velocity at ($H=H_\tindex{c}\,|\,T=0$)
is in general referred to as a continuous phase
transition~\cite{BARABASI_1,NOWAK_1,ROTERS_1}. 
Considering three dimensional systems, for instance, the interface
velocity~$v$  
vanishes at the so-called pinning/depinning transition according to
$v(T=0) \sim (H-H_\tindex{c})^{\beta_{3d}}$
and~$v(H=H_\tindex{c}) \sim T^{1/\delta_{3d}}$, respectively (see
e.g.~\cite{ROTERS_1,AMARAL_1}). 
The correlation length diverges algebraically: $\xi \sim
(H-H_\tindex{c})^{-\nu_{3d}}$~\cite{ROTERS_1,JI_1}.

This pinning/depinning transition can be understood within
the framework of the renormalization group theory
(see e.g.~\cite{WILSON_1,DOMB_1,KADANOFF_1,NATTERMANN_1,CHAUVE_3}
and references therein).
Within this theory it is found that the order parameter of a
continuous phase transition is a generalized homogenous function of,
in general, several thermodynamic parameters~\cite{WILSON_1}.
In many cases temperature and magnetic field belong to these
parameters.
Additional parameters may cause crossovers. 
Examples are spatial anisotropies in
Heisenberg models~\cite{FISHER_4}, dipolar effects~\cite{FISHER_4}, or
restrictions due to geometry like in thin films~(see
e.g.~\cite{CAPEHART_1} and references therein). 
In the case of thin films the system behaves like a three dimensional
system only as long as the correlation length $\xi$ is small as
compared to the film thickness $l$.
Approaching the transition point the component $\xi_l$ of the
correlation length perpendicular to the film layers is bounded by the
film thickness. 
If $\xi_l/l$ becomes of the order of unity a crossover from three to two
dimensional behavior occurs.
Experiments which determine domain wall velocities in magnetic films
typically image the magnetic state of a sample by looking {\it onto}
the sample using the magneto-optical polar Kerr effect and a CCD
camera~\cite{NOWAK_2,LEMERLE_1}.
If the film is sufficiently thin it is possible to obtain the
interface position from snapshots generated by the CCD camera and to 
calculate the interface velocity from the time dependence of this
position. 
This was done in~\cite{LEMERLE_1} where a 
Pt($3.4\,{\rm nm}$)/Co($0.5\,{\rm nm}$)/Pt($6.5\,{\rm nm}$) film with
perpendicular anisotropy was investigated.
Due to the film thickness the authors saw evidence to neglect the
height dependence of the interface position and to treat the interface
not as a two dimensional interface but as a one dimensional line. 
With increasing film thickness, however, this assumption fails because
then 
the correlation length drops below the film thickness causing
a crossover to three dimensional behavior.
This scenario is investigated in the present paper in the context of
driven interfaces in the random-field Ising model.  

\section{simulation}
\label{sec:model_and_sim}
The RFIM is defined by the Hamiltonian 
\begin{equation}
  {\cal H} =
  -\frac{J}{2}\, \sum_{\langle i,j \rangle} S_i \, S_j
  -H \, \sum_{i} S_i
  - \sum_{i} h_i\,S_i
  \; \mbox{.}
\end{equation}
The first term is the exchange interaction 
of neighboring spins and the sum is taken over all pairs of them
($S_i=\pm 1$). 
The second term specifies the coupling to the driving field $H$.
Additionally, the spins are coupled to independent quenched local
random-fields $h_i$ characterized by their 
probability density $p(h_i)$ given by
\begin{equation}
  p(h_i) = 
  \left\{
    \begin{array}{ccl}
      (2\Delta)^{-1} & \;{\rm for}\; &|h_i| < \Delta\\
      0 & & \mbox{otherwise.}
    \end{array}
  \right.
\end{equation}
We use a random-sequential update with transition probabilities
according to a heat bath algorithm in the limit of zero temperature.  
Since in the vicinity of the critical point finite-size effects
may not only occur due to the finite film thickness but also due to the
other finite extensions of our system, we calculated 
the interface velocities for each film thickness $l$ varying the other
extensions of the system. 
For sufficient large extensions we observed no finite-size effects
from which we concluded that the data presented in the following
correspond within negligible errors to those 
of an infinite extended film of thickness $l$. 
Note that for this analysis the extensions of the system must be
increased if one approaches the critical point.
But this increase requires an increase in {\sc CPU}~time which
restricts therefore the data to values not too close to the critical
point. 
We investigate system sizes of up to
$l \times L^2 = 8 \times 1024^2$
unit cells of a body centered cubic lattice.  

We apply periodic boundary conditions in the directions parallel to
the film and antiperiodic ones perpendicular to the interface
(see~\cite{ROTERS_1}). 
The interface moves along the~$[100]$ direction of the
bcc lattice resulting in a finite interface velocity for any driving
field $H \neq 0$ in the absence of disorder. 
The same behavior is found, for instance, on simple cubic lattices
with an interface moving along the diagonal direction of the
lattice~\cite{NOWAK_1,ROTERS_1} and on diamond lattices with the
interface moving along the $[100]$~direction. 
The schematic phase diagram in~\cite{ROTERS_1} applies to
all of these cases.
In particular, a continuous phase transition is found for
$\Delta>J$ as long as nucleation does not occur.
This is the case for $\Delta=3$, which turns out to be a convincing
choice because then the dimensional crossover is numerically
accessible for a broad range of system sizes. 

\section{results}
\label{sec:results}

In our simulations we start with an originally flat interface which is
built into the system.
After a transient regime the interface reaches a stationary
state.
\begin{figure}
  \epsfxsize=8.0cm
 \epsffile{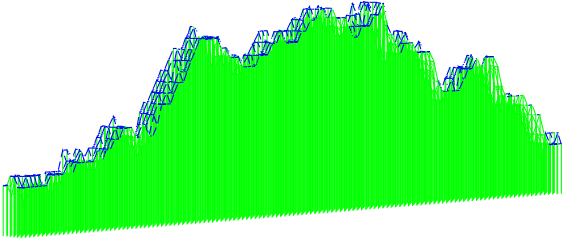}
 \caption{
   Snapshot of a moving interface obtained in a film consisting of 6
   magnetic layers.  
   Overhangs, which are rare and small, are not shown. 
   In the gray sketched area below the interface 
   the spins are aligned parallel to the driving field.
   Spins above the interface are aligned antiparallel and not shown. 
   \label{fig:snapshot}
   } 
\end{figure}
In Fig.~\ref{fig:snapshot} we show the snapshot of an interface
configuration (for $H=1.8$ and $\Delta=3.0$). 
Overhangs are not displayed since they are rare and small. 
We obtain the interface velocity from the mean interface
position~$\bar{h}$ at a given time~$t$ according to
$v=\partial \bar{h}/\partial t$. 
In~\cite{ROTERS_1} it was argued that this velocity
is a generalized homogeneous function of temperature and driving field.
We generalize this ansatz to the present situation and include
additionally a thickness dependence of $v$,
\begin{equation}
  v = \lambda \; v\left(\lambda^{-1/\beta_{3d}} \; \eta ,
  \lambda^{-\delta_{3d}} \; T, \lambda^{-a_l} \; l^{-1} \right)
  \mbox{.}
  \label{eq:ghf2}
\end{equation}
Here, $T$ denotes the temperature and
$\eta=H-H_\tindex{c}^{3d}$ the reduced driving field.
In a bulk system we recover under the assumption
$v(l^{-1}=0)={\rm finite}$ [Eq.~(\ref{eq:ghf2})] 
\begin{math}
  v \sim T^{1/\delta_{3d} } \,
  f\left( \eta \, / \, T^{1/\beta_{3d} \delta_{3d}} \right)
\end{math}
which is known to be satisfied in the RFIM~\cite{NOWAK_1,ROTERS_1}.
In the following we consider only $T=0$. 
Choosing $\lambda=l^{-\beta_{3d}/\nu_{3d}}$ in Eq.~(\ref{eq:ghf2}) we
obtain for the velocity the following scaling behavior: 
\begin{equation}
  v = l^{-\beta_{3d}/\nu_{3d}} f \left( \eta \; l^{1/\nu_{3d}} \right)
  \label{eq:scal}
\end{equation}
with $f(x\to \infty) \sim x^{\beta_{3d}}$.
In this limit the value of $H_\tindex{c}$ coincides with that of the
bulk system.
For any finite $l$, however, $H_\tindex{c}$ becomes $l$ dependent
and is shifted according to 
\begin{math}
  H_\tindex{c}(l) - H_\tindex{c}^{3d} \approx x^\star l^{-1/\nu_{3d}}
\end{math}
(see~\cite{CAPEHART_1} and references therein).
Inserting this relation into Eq.~(\ref{eq:scal}) one finds that $f(x)$
does not vanish at $x=0$ but at a finite value $x^\star$. 

\begin{figure}
 \epsfxsize=8.0cm
 \epsfysize=6.8cm
 \epsffile{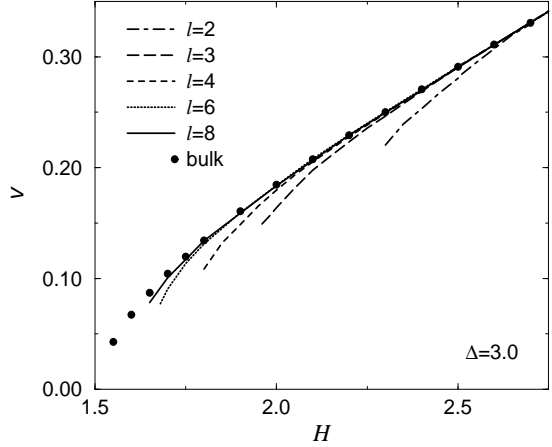}
 \caption{
   Interface velocities $v$ obtained for various film 
   thicknesses~$l$ and different driving fields $H$.
   We also plot the interface velocity of
   the bulk system. 
   Approaching $H_\tindex{c}(l)$ the system size must be increased in 
   order to find negligible finite-size effects
   (see~Sec.~\ref{sec:model_and_sim}).
   The curves terminate at fields where the finite size effects set in
   for the largest set sites accessible numerically.
   \label{fig:unscal_pbc}
   } 
\end{figure}
\begin{figure}
  \epsfxsize=8.0cm
 \epsfysize=6.8cm
 \epsffile{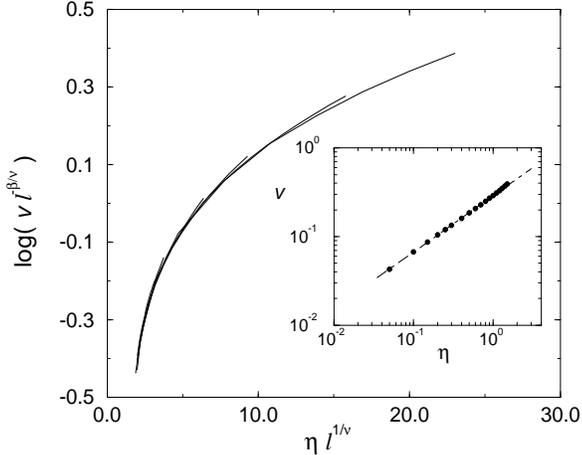}
 \caption{
   Scaling plot according to Eq.~(\ref{eq:scal}). 
   From the data collapse we obtain
   $\beta_{3d}=0.677\pm0.07$, $\nu_{3d}=0.763\pm0.03$ and
   $H_\tindex{c}^{3d}=1.491\pm 0.02$.
   The inset shows the interface velocities of the bulk
   system. We obtain $\beta^{\rm bulk}=0.64\pm0.05$
   and $H_\tindex{c}^{\rm bulk}=1.5\pm0.015$.
   For small values of $\eta$ all curves join 
   the data collapse.
   For sufficient large $\eta$, i.e.\ well above the
   crossover region, the scaling ansatz Eq.~(\ref{eq:scal}) does not
   hold as can be seen from the deviations of single curves from the
   scaling behavior.
   \label{fig:scal}
   } 
\end{figure}

In Fig.~\ref{fig:unscal_pbc} we plot $v(H)$ for different film
thicknesses. 
For a comparison, we also plot velocities obtained in a three
dimensional system.
From the data it is evident that the
curves $v(H)$ for different film thicknesses deviate more and more
from the bulk behavior with decreasing $H$.
In the corresponding region of $H$-values the crossover from 2 to
3-dimensional behavior occurs which thus is numerically accessible.
In the crossover region the interface velocity in a film turns out to 
be smaller than in the bulk meaning that the threshold field
$H_\tindex{c}$ is shifted towards {\it larger} values of the driving
field. 
Rescaling the velocities according to Eq.~(\ref{eq:scal}) yields the
data collapse shown in Fig.~\ref{fig:scal}.
From the collapse it may be concluded that the scaling function
$f(x)$~[see Eq.~(\ref{eq:scal})] 
vanishes at a value $x^\star>0$,
again meaning that the threshold field
is shifted towards larger values with decreasing
film thickness.
We obtain $\beta_{3d}=0.677\pm0.07$
and $H_\tindex{c}^{3d}=1.491\pm0.02$.
These values coincide within the error-bars with those found in bulk
systems. 
In this case (inset of Fig.~\ref{fig:scal}) we find
$\beta^{\rm bulk}=0.64\pm0.05$ and
$H_\tindex{c}^{\rm bulk}=1.5\pm0.015$ 
confirming the results of the crossover scaling. 
From the data collapse in Fig.~\ref{fig:scal} we also obtain 
$\nu_{3d}=0.763\pm0.03$.
Both $\beta_{3d}$ and $\nu_{3d}$ coincide within the error-bars
with values found on other lattices and different values of
$\Delta$~\cite{ROTERS_1,AMARAL_1,JI_1}. 
Also, they agree with the exponents of the Edwards-Wilkinson equation
with quenched disorder (see e.g.~\cite{NATTERMANN_1,CHAUVE_3,EDWARDS_1} 
and references therein).
For this equation 
$\beta_\tindex{QEW}\approx 0.62$ and
$\nu_\tindex{QEW}\approx 0.77$ was obtained by an
$\epsilon$-expansion valid to order~$\epsilon^2$~\cite{CHAUVE_3}.

Since the interface velocity satisfies the ansatz Eq.~(\ref{eq:scal})
it is possible to draw conclusions about the critical behavior at
$H_\tindex{c}(l)$.
Considering $f(x)$ with $x=\eta \, l^{1/\nu_{3d}}$ and taking into account
that $\nu_{3d}>0$ one finds $x=0$ at $H_\tindex{c}(l)$ for any finite
film thickness.
On the other hand, a three dimensional system corresponds to the limit
$l\to\infty$ and in this case $f(x)$ is determined by the limiting
behavior $f(x\to\infty)\sim x^{\beta_{3d}}$.
The critical behavior in a film is therefore different as compared to
the bulk behavior and the critical exponents $\beta$ and $\nu$ coincide
with those of the two dimensional model for any finite film
thickness.
In the two dimensional model $\beta_{2d}\approx 0.31$ and
$\nu_{2d} \approx 1$ was found~\cite{NOWAK_1,AMARAL_1}.
Taking Eq.~(\ref{eq:ghf2}) into account, we also conclude that 
the exponent $\delta$ characterizing the thermal rounding of the
depinning transition
[$v \sim T^{1/\delta_{2d}}$ at $H=H_\tindex{c}(l)$] is given by
$\delta_{2d} \approx 5$ (see~\cite{NOWAK_1}) for any finite film thickness.

\section{conclusion}
In conclusion, we have investigated the depinning transition of a
driven interface in thin films.
We have found that the critical behavior is governed by the two
dimensional fix-point and the corresponding exponents
for any finite film thickness.
The exponents obtained by the crossover scaling are in
agreement with those of the Edwards-Wilkinson universality class. 
The scaling ansatz used to analyse our data could also be used to
determine critical exponents of the 
depinning transition of three dimensional systems experimentally, in
particular since present experimental techniques use thin films as
samples~\cite{NOWAK_2,LEMERLE_1}.  
In~\cite{NOWAK_2}, for instance, ${\rm Co}_{28}{\rm Pt}_{72}$ alloy
films were investigated with grain sizes of typically $20\,{\rm nm}$
and film thicknesses of $5...50\,{\rm nm}$. 
If one naively assumes that by a variation of temperature and/or
driving field it is possible to increase the correlation length from
the size of the grain to the film thickness we expect the crossover
scaling [Eq.~(\ref{eq:ghf2})] 
to work and to yield the exponents of a three dimensional sample. 
Note that due to dipolar interactions these exponents need not
to coincide with those of the RFIM.

\acknowledgments
The authors would like to thank S.\ L\"ubeck for many useful
discussions. 
This work was supported by the Deutsche Forschungsgemeinschaft via
GRK~277
{\it Struktur und Dynamik heterogener Systeme} 
(University of Duisburg)
and SFB~491
{\it Mag\-ne\-tische Hetero\-schich\-ten: Struktur und el\-ek\-tronischer
  Transport}
(Duisburg/Bochum).


\end{document}